\documentclass[aps,showpacs,twocolumn,eqsecnum,floatfix]{revtex4}

\usepackage{epsfig}
\usepackage{latexsym}
\usepackage{mathbbol}

\begin{document}

%%%%%%%%%%%%%%%%%
%%%   TITLE   %%%
%%%%%%%%%%%%%%%%%

\title{On a debate about cosmic censor violation}

\author{Miguel Alcubierre}
\email{malcubi@nuclecu.unam.mx}

\author{Jos\'e A. Gonz\'alez}
\email{cervera@nuclecu.unam.mx}

\author{Marcelo Salgado}
\email{marcelo@nuclecu.unam.mx}

\author{Daniel Sudarsky}
\email{sudarsky@nuclecu.unam.mx}

\affiliation{Instituto de Ciencias Nucleares, Universidad Nacional
Aut\'onoma de M\'exico, A.P. 70-543, M\'exico D.F. 04510, M\'exico.}

%%%%%%%%%%%%%%%%
%%%   DATE   %%%
%%%%%%%%%%%%%%%%

\date{\today}

%%%%%%%%%%%%%%%%%%%%
%%%   ABSTRACT   %%%
%%%%%%%%%%%%%%%%%%%%

\begin{abstract}

We review the arguments and counter arguments about the recent proposal for generic censorship 
violation. In particular the argument made in~\cite{His counterargument} against our proposal 
for a possible expanding domain wall
that could encompass a large black hole, is shown to have a serious flow. 
Other problems of the original idea are also discussed.

\end{abstract}

%%%%%%%%%%%%%%%%
%%%   PACS   %%%
%%%%%%%%%%%%%%%%


%%%%%%%%%%%%%%%%%%%%%
%%%   MAKETITLE   %%%
%%%%%%%%%%%%%%%%%%%%%

\maketitle

%%%%%%%%%%%%%%%%%%%%%
%%%   THE MODEL   %%%
%%%%%%%%%%%%%%%%%%%%%

\section{The model}

There has been a recent upsurge in interest in the possibility of a generic violation 
of Cosmic Censorship, motivated by the arguments put forward in~\cite{Horowitz1}.
This work describes a type of situation that is argued would lead to evolution from  
smooth initial data to a naked singularity for certain type of scalar field models with 
asymptotic Anti de Sitter (AdS) space-times.
Specifically the model considers a scalar field minimally coupled to gravity and 
having a self interaction potential with two local minima having negative values: 
$V(0) = -3 V_0$ and $V(\phi_1) = -3 V_1$ with $0<V_1<V_0$. The potential is chosen
to have a small positive barrier between these minima and is required to satisfy the 
positive energy condition among the configurations that are asymptotically  AdS 
corresponding to the false vacuum $\phi_1$ (i.e. the effective cosmological constant 
being $\Lambda_{\rm eff} = -V_1$ in units where $8 \pi G =1$). That is, we consider  
initial data consisting of the three metric $\gamma_{ab}$, 
the extrinsic curvature $K_{ab}$, the scalar field $\phi$ and its conjugate momentum $\Pi$, 
which satisfy the Hamiltonian and momentum constraints and 
which correspond to an asymptotically AdS geometry.

Within the spherical symmetric situation one chooses  to parametrize the metric as
\begin{equation}
 d\sigma^2 = \left( 1- {2m(r)\over r} +V_1 r^2 \right)^{-1} dr^2 + r^2 d\Omega^2
\end{equation}
 
The first thing we must note is that such parametrization is appropriate as long as 
no ``bag of gold" type configuration needs to be considered.  
For any configuration having the asymptotic parametrization given above one uses the 
notion of ``ADM" mass given  by

\begin{equation}
M_{\rm ADM} = \lim_{r \rightarrow \infty} m(r)
\end{equation}

One considers now the set of initial data $C[R_1] = \{${\it the set of configurations 
that correspond to} $\phi=\phi_1$ {\it for all} $r >R_1$ {\it and} $\phi(0)=0 \}$.
Within this set one considers the instantaneously static subset  consisting of 
configurations having $K_{ab} = 0$ and $\Pi=0$.
The mass of such a configuration can be expressed by use of the Hamiltonian constraint as:
\begin{equation}
M_{\rm ADM} = \frac{V_1}{2} R_1^3 + \tilde M_V R_1^3 + \tilde M_\partial R_1
\label{eq:M_ADM}
\end{equation}
where  $M_V$ and $M_\partial$ are expressions (the first related to the contribution of 
the scalar field potential and the second related only to the scalar field gradients) 
that depend only on the scalar field configuration expressed in terms of a rescaled radial 
coordinate $y=r/R_1$ and thus do not depend explicitly on $R_1$.
Concretely the functionals are given by
\begin{eqnarray} 
\tilde M_V[\phi(y)]&\equiv&\frac{1}{2}\int_0^1 e^{-\int_y^1 \frac{\hat{y}}{2}(\partial_y \phi)^2 d\hat{y}}
V(\phi) y^2 dy \nonumber \\ 
\tilde M_{\partial}[\phi(y)]&\equiv&\frac{1}{2}\int_0^1 e^{-\int_y^1 \frac{\hat{y}}{2}(\partial_y \phi)^2 d\hat{y}}
\frac{1}{2} (\partial_y \phi)^2 y^2 dy \nonumber
\end{eqnarray}
Thus, if one keeps the rescaled configuration fixed (i.e. one keeps $\phi(y) $ fixed) the 
scaling properties of the mass are given by the expression~(\ref{eq:M_ADM}).
The situation considered in~\cite{Horowitz1} involves adjusting the  parameters of the
theory so that for the configuration that minimizes $\tilde M_V $ (denoted by $\phi_0(y)$) the terms 
proportional to $R_1^3$ cancel out, i.e. $V_1 = -2 \tilde M_V [\phi_0(y)] $.
The argument for cosmic censorship violation is the following:  
The configuration $\phi_0(y)$ has a value $y_0<1$ such that for $y<y_0$, the configuration 
is very close to the true vacuum $\phi=0$, i.e. $\phi_0(y) <\epsilon$ for a small epsilon.
The corresponding region is essentially a region of radius $r_0=R_1y_0$ of AdS spacetime 
with a cosmological constant given by $V_0$ and a scalar field slightly removed from 
the true minimum. Such a configuration is known to evolve to a singularity.
The ADM mass of this configuration is $M_0(R_1) = \tilde M_\partial [\phi_0] R_1$.
Let's assume that a black hole develops and that it has at asymptotically late times a 
radius $R_{BH}$. Assuming it is a standard AdS black hole corresponding to the asymptotic 
value of the effective cosmological constant, its ADM mass is given by 
$M_{BH}=(1/2)( R_{BH} + V_1R_{BH}^3)$.
Its radius therefore must satisfy  $(1/2)(R_{BH}+ V_1R_{BH}^3) < M_0(R_1)$. In particular 
$R_{BH} < (2M_0(R_1)/ V_1)^{1/3} = C R_1^{1/3}$.
As the matter fields in this theory satisfy the null energy condition, the area of the event 
horizon has to be an increasing quantity in the sense that given two Cauchy hypersurfaces 
the intersection of the event  horizon with the later one has larger area than its 
intersection with the earlier one. Thus the intersection of the initial data hypersurface 
with the event horizon has to occur (if at all) at $r_{intersection}< R_{BH} <C R_1^{1/3}$. 
Clearly one can choose $R_1$ sufficiently large so that $r_0=R_1y_0$ be much larger than 
$C R_1^{1/3}$ given the fact that the former scales like $R_1$ while the latter scales like 
$R_1^{1/3}$.
In this situation, it is argued that the black hole can not encompass the region that evolves 
into a singularity. Thus the singularity that results from the evolution should be naked.

%%%%%%%%%%%%%%%%%%%%%
%%%   DISCUSION   %%%
%%%%%%%%%%%%%%%%%%%%%

\section{Discussion}

The argument presented by us in~\cite{Our Comment} suggests the possibility that given the initial
conditions set up in~\cite {Horowitz1}, a type of domain wall (connecting  the local and global
minima of the potential) will expand continuously, leading to a condition where  a large black
hole could form in the interior region (the one corresponding to the global minimum),
In~\cite{His counterargument} it is argued that ``{\it It is easy to show that in our case 
this could not occur" because``the region where $\phi$ is close to the global minimum (of the 
potential) cannot expand without increasing the total energy. This is because the initial profile
$\phi(y)$ is already chosen to minimize the potential contribution to the energy... Any other 
shape for the wall will have higher energy}". 
There is a serious flaw in this argument: The initial profile was chosen to minimize the value 
of $\tilde M_V$, and {\bf not that of the total energy: $M_{\rm ADM}$}. In fact the configuration 
$\phi_0(y)$ could in principle have a total energy that is quite higher that the minimum within  
$C[R_1]$. If one were to choose as initial configuration the true minimum of the total energy  
in $C[R_1]$, the scaling properties (as one changes the value of $R_1$) would certainly differ 
from those~(\ref{eq:M_ADM}) above.
Thus, there is a clear possibility, as suggested in~\cite{Our Comment}, that the configuration will evolve  
into a barrier that expands indefinitely (i.e. $R_1$ would expand), while the energy at every
``instant" of the corresponding frozen configuration (i.e. one where the kinetic terms are made 
zero by hand) decreases, the term $\tilde M_V$ increasing slowly and the term $\tilde M_\partial$ 
decreasing faster (as needed). The true total energy of the configuration ($M_{\rm ADM}^{\rm true}$) 
being of course conserved, after adding the kinetic terms in the actual solution 
($M_{\rm ADM}^{\rm true} = M_{\rm ADM}^{\rm frozen} + M^{\rm kinetic}$).
What is known is therefore that $\tilde M_V$ has to increase relative to its initial value and that 
$\tilde M_\partial$ has to be positive definite. 
The issue is then if one can envision an evolution (with some appropriate time parameter $t$ 
labeling the spacetime foliation) such that the instantaneous configuration $\phi(r,t)$  differs 
from $\phi_1$ only for values of $r<R_1(t)$, and where $R_1(t)$ increases without bound.
Energetically all we need is to show that it is possible for the corresponding ADM mass of the 
instantaneously frozen configuration to decrease.
It is easy to construct an example compatible with what is known about the functionals $\tilde M_V$ 
and $\tilde M_\partial$:
Let us take as parameter the value of $ R=R_1(t)$ rather than $t$. So the instantaneous configuration 
will be described by $\phi(y,R)$.
As a result, through the dependence of the configuration on $ R$, the functionals become, when 
evaluated on $\phi(y,R)$, functions of $R$. About these functions we know, in principle, only 
that $ \tilde M_V [R]$ increases relative to its value at $R_0=R_1(t=0)$, and that 
$\tilde M_\partial [R]$ is positive definite.
Now one can give a simple example of functions satisfying the required behaviour:
\begin{equation}
\tilde M_V [R] = \tilde M_V [R_0]  + A (R-R_0){R^{-9/2}} \label{eq:m_pot}
\end{equation}
where $A=\tilde M_\partial [R_0] R_0^{3/2}$, and
\begin{equation}
\tilde M_\partial [R]= \tilde M_\partial [R_0]  \left( \frac{R_0}{R}
\right)^{5/2}
\end{equation}
In this way
\begin{equation}
M_{\rm ADM}^{\rm frozen}[R]= V_1 R^3 + \tilde M_V R^3 + \tilde M_\partial R =
\tilde M_\partial(R_0) \frac{R_0^{3/2}}{R^{1/2}}
\end{equation}
which is clearly a decreasing function of $R$. As mentioned above the true mass of the evolving 
configuration will be increased relative to the value above by the kinetic terms, leading to a 
mass that is conserved through out the evolution. Therefore, and in contrast with the claims made
in~\cite{His counterargument} and seconded by~\cite{Hubeny}, these energetic arguments can not be used to exclude the 
possibility that a domain wall will developed in the situation that has  been proposed 
in~\cite{Horowitz1}. 
Furthermore, as the arguments in~\cite{Horowitz1} are supposed to refer to a generic situation
(i.e. they are presented as evidence of a generic ``violation of Cosmic Censorship"), the
discussion of the issue at hand can not (if the generic nature of the argument is to be preserved)  
be based on the detailed properties of a specific configuration (i.e. the absolute minimum of  
$\tilde M_V$).

In a recent work~\cite{His counterargument}, the original proponents of the generic cosmic censor 
violation have argued that the connection between the almost homogeneous region and a singularity
is in doubt due to possible influences of the arbitrarily far away regions in finite time in AdS 
spacetimes. 

One final issue that needs to be clarified is the following, even if one knew that the starting 
configuration had no ``bag of gold" present, a situation that would clearly invalidate the 
initial parametrization of the metric, there is the possibility that a ``bag of gold" 
configuration might appear as a result of the evolution. In such a case all the region 
containing the singularity could be arbitrarily large and still be contained  within a black 
hole of arbitrary small area, completely circumventing the arguments of~\cite{Horowitz1}.
In fact, even in the standard collapse of a spherical distribution of dust leading to the 
formation of a Schwarzschild black hole, one con encounter the formation of such ``bags of 
gold": The vacuum region would eventually include points within the event horizon, which, 
by virtue of Birkoff's theorem, must correspond to the extended Schwarzschild metric. Within the
event horizon the surfaces of constant value of  $r$ (the radial area parameter of standard 
Schwarzschild coordinates) are spacelike, thus a given Cauchy hypersurface can easily  cross 
them in both directions. In such situation the Cauchy surfaces would have a bag of gold type
of structure: The value of the area of the orbits of isotropic symmetry, would, as we move inwards,  
change from being a decreasing quantity into an increasing one.

%%%%%%%%%%%%%%%%%%%%%%
%%%   CONCLUSION   %%%
%%%%%%%%%%%%%%%%%%%%%%

\section{Conclusion}
\label{sec:conclusion}

In conclusion, there are at this point various issues that cast doubts on the argument of~\cite{Horowitz1}: 
The one raised in~\cite{Our Comment} and discussed here in more detail, and the one raised 
in~\cite{His counterargument}, and the issue regarding the possibility of the generation of bag of 
gold. However the main issue that is raised as the result of~\cite{Horowitz1}: whether or not a Big Crunch
singularity~\cite{Dafermos} arises in the {\it original} proposed situation, remains at this time an important and open 
question.

%%%%%%%%%%%%%%%%%%%%%%%%%%%
%%%   ACKNOWLEDGMENTS   %%%
%%%%%%%%%%%%%%%%%%%%%%%%%%%

\section{Acknowledgments}
This work was supported in part by \linebreak CONACyT through grant 149945, by
DGAPA-UNAM through grants IN112401, IN122002 and IN108103-3, and by DGEP-UNAM
through a complementary grant.

%%%%%%%%%%%%%%%%%%%%%%
%%%   REFERENCES   %%%
%%%%%%%%%%%%%%%%%%%%%%

%\bibliographystyle{bibtex/apsrev}
%\bibliography{bibtex/referencias}

\begin{thebibliography}{99}

\bibitem{Horowitz1} T. Hertog, G. T. Horowitz and K. Maeda. Phys. Rev. Lett. {\bf 92} 131101 (2004). gr-qc/0307102 
\bibitem{Our Comment} M. Alcubierre, J. A. Gonzalez, M. Salgado and D. Sudarsky. gr-qc/0402045
\bibitem{His counterargument} T. Hertog, G. T. Horowitz and K. Maeda. gr-qc/0405050 
\bibitem{Hubeny} V. E. Hubeny, X. Liu, M. Rangamani and S. Shenker. hep-th/0403198 
\bibitem{Dafermos} M. Dafermos. gr-qc/0403033 
 
\end{thebibliography}

%%%%%%%%%%%%%%%
%%%   END   %%%
%%%%%%%%%%%%%%%

\end{document}